\documentclass[reprint]{revtex4-2}

\usepackage{graphicx}
\usepackage{dcolumn}
\usepackage{bm}
\usepackage{braket}
\usepackage{chemformula}
\begin{document}

\preprint{APS/123-QED}

\title{A brief perspective of high temperature superconductivity in the cuprates: \\ Strong correlations combined with superexchange match experiment}

\author{J\'ozef Spa\l{}ek}
 \affiliation{Institute of Theoretical Physics, Jagiellonian University, ul. \L{}ojasiewicza 11, PL-30-348 Krak\'{o}w, Poland}
 \email{jozef.spalek@uj.edu.pl}

\date{\today}

\begin{abstract}
High temperature superconductivity encompasses the cuprates, nickelates, iron pnictides, and LaH$_x$ compounds. The first three groups of compounds involve in the pairing electrons, which are strongly to moderately correlated, whereas in the last class of systems specific phonon excitations. In this overview we concentrate first on the (semi)quantitative theory of high T$_{C}$ superconductivity in the cuprates based on our original vibrational approach beyond the renormalized mean field theory. The model we explore mainly is $t$-$J$-$U$ model containing both the superexchange (kinetic energy) combined with strong interelectronic correlations. Selected equilibrium and dynamic-excitation properties are analyzed briefly. General questions regarding the pseudogap and two--dimensional character of those systems are raised. 
\end{abstract}

\maketitle


\section{Introduction}

Originally, by high--temperature (high--T$_{C}$) superconductors we understood that discovered 
35 years ago in La$_{x-1}$B$_{x}$TiO$_{y}$, La$_{x-1}$Sr$_{x}$CuO$_{y}$, 
and YBa$_{2}$Cu$_{3}$O$_{7-\delta}$ systems. Later, the iron pnictide and chalogenide
systems such as LaFeAsO$_{1-x}$F$_{x}$ and FeSe. Recently, the 
nickelates LaNiO$_{2}$ and related compounds have been studied intensively. A separate class is formed by the LaH$_{10+x}$, for which the critical temperature has 
reached 250 K or even higher value. The principal difference between the 
hydrogen--rich and remaining system is that in the case of LaH$_{10+x}$ the pairing of 
electrons seems to be caused by phonons, whereas in the cuprates, nickelates, and iron pnictides the 
strong to moderate interelectronic correlations play a decisive role. The aim of this brief 
overview is to compare our theoretical results for the cuprates with the principal experimental results
in a consistent and quantitative way. 

The structure of this brief review is as follows. In Sec. II we discuss principal characteristics of the cuprates. In Sec. III we overview the qualitative features of our theory, whereas in Sec. IV we provide explicit examples of a quantitative comparison of our results with experimental data. A brief outlook is deferred to Sec. V. This paper aims to specify and summarize the most important results elaborated in a recent topical review \cite{Spalek_phys_rep}. 

\section{Principal characteristics of the cuprates}

\subsection{Structural and electronic specific features of the cuprates}

The most striking structural property of the high--temperature superconducting cuprates and pnictides 
is their quasi--two--dimensionality, composed in the simplest situation of well separated 
CuO$_{2}$ planes. This is the case for e.g. 
La$_{{1}-\delta}$Sr$_{\delta}$CuO$_{{4}}$ or 
Bi$_{2}$Sr$_{1.6}$La$_{0.4}$CuO$_{{6}-\delta}$ mixed compounds. This simplifying assumption induced series of 
studies of strictly two--dimensional models of high-$T_{C}$ superconductivity, even though 
it is not exactly clear whether, strictly speaking, a spatially homogeneous two--dimensional 
transition to superconducting state is possible at nonzero temperature ($T>0$). The evidence that 
such an ordering is supported by the results for single-plane of FeSe \cite{Wang2017}. 

\begin{figure*}
    \centering
    \includegraphics[width=0.8\textwidth]{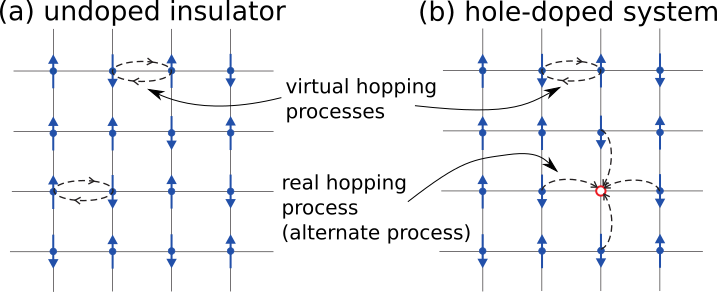}
    \caption{Schematic representation of the particle dynamics in terms of hopping processes (dashed arrows) in the Mott-insulating state (a) and the
strongly-correlated metal phase (b). Virtual hopping involves two consecutive direct hopping processes back and forth and occurs in both cases, (a) and (b). The direct
hopping results in real motion of holes and occurs only in the strongly-correlated metal phase (b). In the strong-correlation regime, the direct hopping
processes via doubly occupied configurations $\ket{\uparrow\downarrow}$ are precluded. In the last case we speak about extreme strong correlations. The arrows surrounding the hole (red circle) mark possible real hoppings around it.}
 \label{fig:my_label}
\end{figure*}

The second most important feature is that the carriers in CuO$_{2}$ plane are holes in the 
doped Mott insulator. The situation is schematically 
depicted in Fig. \ref{fig:my_label}. Note that only Cu$^{2+}$ ions are shown for the sake clarity. Virtual 
hopping processes, specified also there, lead to the antiferromagnetic kinetic exchange representing the superexchange, whereas the real hopping processes provide the charge transport of single carriers and 
their pairing in both hole-- and electron--doped cases situations. Note a gradual character of transformation from antiferromagnetic Mott insulator to a strongly correlated metal with doping. At this point, it is fair to say that so far it is not clear whether this changeover from the Mott insulator to the strongly correlated metal is a real quantum phase transition with an incipient quantum critical point, blurred by the substitutional disorder (e.g. Sr for La), taking place in the  La--Sr--O insulating planes, sandwiching the periodic arrangement of the CuO$_{2}$ planes, where the actions goes.

\begin{figure*}
    \centering
    \includegraphics[width=0.8\textwidth]{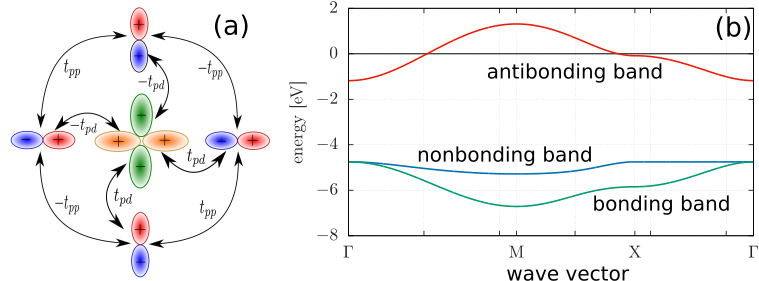}
    \caption{Bare (without interaction) three--band structure in the tight binding approximation. (a) Definition of the hopping parameters between the p$_{x}$ , p$_{y}$ , and d$_{{x}^2 -{y}^2}$ orbitals, with the sign convention for the antibonding orbital
structure. This structural unit forms a basis of three-band model of the CuO$_{2}$ plane in the cuprates. (b) The band structure of the $d$-$p$ model with microscopic parameters: $t_{pd} \sim 1.13\text{ }eV$, $t_{pp} \sim 0.49\text{ }eV$, and $\epsilon_{pd} \sim 3.57\text{ }eV$. The Fermi energy is taken as the reference value and
corresponds to the filling $n = 5$ per Cu$^{{2+}}$O$_{2}^{2-}$ complex, corresponding to half--filled antibonding band. This partially filled band is split off by about $\epsilon_{pd}$ from the remaining filled bands and reflects a single--hybridized (bare) band, the horizontal line marks the position of Fermi energy for $n=5$ electrons per CuO$_{2}$ unit (after \cite{Wang2017}). In the strong correlation limit the antibonding band (and the other two) is split into two Hubbard subbands.}
    \label{fig:fig2}
\end{figure*}

Another striking feature is the circumstance that the CuO$_{2}$ planes may be represented 
originally by the atomic 3d$_{{x}^2-{y}^2}$ states, representing the highest positioned electron of 
nominally 3d$^{9}$ shell of Cu$^{2+}$ ion, hybridized with two 2p$_{x}$ and 2p$_{y}$ 
states of nominally O$^{2-}$ ions. The situation is 
depicted schematically in Fig. \ref{fig:fig2} (left). This 3--orbital periodic structure, arranged into a square lattice, leads to the 
three bands specified in Fig. \ref{fig:fig2}b; the t$_{\alpha\beta}$ parameters are the hopping integrals
between the specified orbitals. 
Now, if the CuO$_{2}$ system is regarded as effectively single--band system composed of 
3d$_{{x}^2-{y}^2}$ orbitals dressed with 2p$_{x,y}$ orbitals, in which the latter 
states play only a passive role \cite{Zegrodnik_2019_PRB}, then such a single--band Mott insulator is represented by a singly occupied set of Cu$^{{2+}}$ ions. This is the situation depicted in Fig. \ref{fig:my_label}, 
where the $\uparrow$ and $\downarrow$ arrows specify the spins of the ninth electrons per site, 
may be regarded as the model situation of the Mott--Hubbard insulator. The doping $\delta \equiv 1-n$ 
represents then the average number of hole carriers counted per site. If however, one takes the 
three-orbital 3d$_{{x}^2-{y}^2}$--2p$_{x,y}$ model, then the corresponding Mott-Hubbard insulator (called in that case the charge--transfer insulator) contains 5 electrons per Cu$^{2+}$O$^{2-}_{2}$ cluster (two 2p electrons per oxygen and one electrons per copper). In that situation the hole doping may be defined as $\delta \equiv 5-n$, where $n$ in both situations is the number of electrons per fundamental unit, Cu or CuO$_{2}$, respectively. 
The fourth principal assumption is that the original microscopic parameters such as the hoppings $t_{\alpha\beta}$ or interaction strength do not vary essentially in the whole doping range where superconductivity appear, i.e., for $0 	\lesssim \delta \lesssim 1/3$. Those bare parameters do vary from system to system, but mainly due to interelectronic correlations which induce their strong doping dependent renormalization. They complement the bare one--electron structure. Effectively, one should regard the single--band model description as that referring to the situation of the antibonding band (cf. Fig. \ref{fig:fig2}b), containing effectively $1-\delta$ electrons per copper. This point is to be verified later. 

The final structural feature of the system is the experimental observation that the electronic properties in the normal state are those of practically two--dimensional metal, with the resistivity in plane/across plane $\rho_{\parallel}/\rho_{\perp} \sim 10^5$ in the optimal situation, and with metallic/semiconducting behavior of $\rho_{\parallel}/\rho_{\perp}$, respectively. On the contrary, the superconducting phase is three--dimensional. This means that there is $d=2$ to $d=3$ dimensional changeover at the critical temperature. In other words, the interplanar coherence appears in the condensed state. This can be clearly shown when examining systematically single-- versus multi--planar systems critical temperature T$_{C}$ as a function of the number of closely spaced planes \cite{Byczuk_PRB}. 

\subsection{Theoretical models of strongly correlated square planar structure of the cuprates}

In our group we have concentrated on studying two theoretical models: The (extended) single--band Hubbard model under the acronym of t--U--J--V model (cf, Appendix A), as well as on three--band 3d--2p$_{{x,y}}$ model. In the latter situation also its similarity to the one--band case under special circumstances has been explored \cite{Zegrodnik_2019_PRB}. The first of them represents the most general single--band model with short--range intersite interactions (and correlations), which reduces to either $t$-$J$ or Hubbard model in proper limits. The three--band model, in turn, allows for an explicit discussion of the role of oxygen in the particle dynamics and ordering, particularly in the metallic state. We overview each of them separately in the context of concrete results and compare them with experiment.

\section{The method and its qualitative interpretation}

\subsection{The method: Single--band model}

The most general single--band model of correlated electrons has been discussed briefly in Appendix A. In this section we limit ourselves to the so--called $t$-$J$-$U$-V model in the form \cite{Spalek_PRB_2017}

  \begin{align}
    \label{hamiltonian}
    &\hat{\mathcal{H}} = \sideset{}{'}\sum_{ij\sigma} t_{ij} \hat{a}_{i\sigma}^\dagger \hat{a}_{j\sigma} + U \sum_i \hat{n}_{i\uparrow} \hat{n}_{i\downarrow} +  \sideset{}{'}\sum_{ij} J_{ij} \hat{\mathbf{S}}_i \hat{\mathbf{S}}_j  \\ \nonumber &+ \frac{1}{2}\sideset{}{'}\sum_{ij} \left(V_{ij} - \frac{1}{2} J_{ij}\right) \hat{n}_i \hat{n}_j.
  \end{align}

\noindent
\begin{figure*}
    \centering
    \includegraphics[width=1\textwidth]{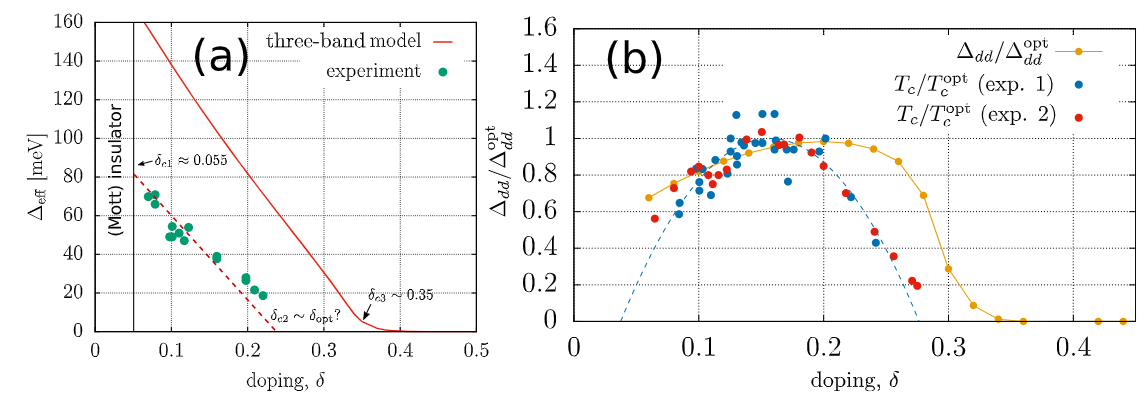}
    \caption{(a) Plot of the experimentally observed pseudogap (points) as compared to theoretical results for the effective single--particle gap $\Delta_{eff}$ obtained within the three--band model. The Hamiltonian parameters are: $U_d = 11 \text{ }eV$, $U_p = 4.1 \text{ }eV$, $\epsilon_{dp} = 3.2 \text{ }eV$, $t_{pp} = 11 \text{ }eV$. The critical doping levels and the Mott--insulator boundary is also marked. (b) Relative correlated $d$-wave gap component with intersite Coulomb interaction of magnitude $V_{dd} = 0.7 \text{ }eV$. Experimental data sets 1 and 2 are taken from Ref. \cite{Huffner_2008}. For a brief discussion of the role of quantum fluctuations in bringing the theoretical results to those obtained from experiment see Outlook. }
    \label{Fig3}
\end{figure*}
The parameters and consecutive terms are defined and explained in Appendix A. Hamiltonian \eqref{hamiltonian} is used as a starting point for a further analysis and solution fro many--particle states. In our comprehensive review \cite{Spalek_phys_rep} we selected the approach based on a trial variational wave function and subsequently have constructed a systematic diagrammatic expansion (DE--GWF) which in the lowest order, provides renormalized mean--field theory (RMFT) in the form of statistically consistent Gutzwiller approximation (SGA) \cite{Jedrak_PRB}. In general, the approach is based on selection of the ground--state many--particle wave function $\ket{\psi_G}$ in the form

\begin{align}
    \ket{\psi_G} \equiv \hat{P}\ket{\psi_0}, 
    \label{wav}
\end{align}

\noindent
where $\ket{\psi_0}$ represents an uncorrelated (single--particle) state, to be defined later in the process of solving the model in a self-consistent manner. The nontrivial projection operator $\hat{P}$ is given by \cite{Bunemann_2012}

\begin{align}
    \hat{P} \equiv \prod_i \hat{P}_i = \prod_i \lambda_{i\Gamma} \ket{\Gamma}_{i} {}_{i}\bra{\Gamma}
\end{align}

\noindent
with the wave--function variational parameters $\lambda_{i,\Gamma} \in \{\lambda_{i,0},\lambda_{i,\uparrow},\lambda_{i,\downarrow},\lambda_{i,\uparrow\downarrow}\}$, corresponding to the local (lattice site \textit{i}) states $\ket{\Phi}$, $\ket{\uparrow}_i$, $\ket{\downarrow}_i$, and $\ket{\uparrow\downarrow}_i$, respectively. The consecutive states represent the empty, single occupied with spin quantum number $\uparrow$ and $\downarrow$, and doubly--occupied states, all on site $i$. For such a choice of the site representation, the $\lambda_{i,\Gamma}$ parameters weight the relative probability amplitudes of local occupancies appearance for each site. In the limit of large Coulomb repulsion ($U \gg W$, where W is the bare bandwidth) the double occupancies are absent. Additionally, we consider here translationally invariant paramagnetic state for which $\lambda_{i\uparrow} = \lambda_{i\downarrow} = \lambda_{i}$. 

The ground state energy is determined by minimizing the variational expression for the ground state energy

\begin{equation}
 E_G \equiv \langle\mathcal{\hat{H}} \rangle_G=\frac{\langle\Psi_G|\mathcal{\hat{H}}|\Psi_G \rangle}{\langle\Psi_G|\Psi_G \rangle}=\frac{\langle\Psi_0|\hat{P}\mathcal{\hat{H}}\hat{P}|\Psi_0 \rangle}{\langle\Psi_0|\hat{P}^2|\Psi_0 \rangle},
 \label{var_ham}
\end{equation}

It turns out that by introducing the following additional ansatz for the $\hat{P}_i$ operator \cite{Bunemann_2012}

\begin{align}
    \hat{P}^2_i \equiv 1 + xd^{HF}_i, 
\end{align}

\noindent
where $x$ is yet another variational parameter and by defining the quantities

\begin{align}
    d^{HF}_i \equiv \hat{n}^{HF}_{i\uparrow}\hat{n}^{HF}_{i\downarrow}, \text{  }\hat{n}^{HF}_{i\sigma} \equiv \hat{n}_{i\sigma} - \braket{\hat{n}_{i\sigma}} \equiv \hat{n}_{i\sigma} - n_0, 
\end{align}

\noindent
with $n_0 \equiv \braket{\psi_0|\hat{n}_{i\sigma}|\psi_0}$, we can perform a systematic expansion of the functioanl \eqref{var_ham} (for details see \cite{Spalek_phys_rep, Zegrodnik_2019_PRB}) and obtain explicitly the interesting us physical properties in the correlated state which are determined thorough corresponding quantities in uncorrelated state. Before detailed physical discussion we should mention the method of defining the uncorrelated wave function $\psi_0$. Namely, it is determined from another variational principle \cite{Kaczmarczyk_Phil_Mag}

\begin{align}
    \frac{\delta}{\delta \bra{\psi_0}} \{\mathcal{F} - \lambda(\braket{\psi_0|\psi_0}-1)\} = 0, 
    \label{var}
\end{align}

\noindent
where $\mathcal{F} \equiv \braket{\hat{\mathcal{H}}}_G$ expressed in terms of uncorrelated correlation functions; here there two intersite functions

\begin{align}
    P_{ij} \equiv \braket{\hat{c}^{\dagger}_{i\sigma}\hat{c}_{j\sigma}}, \text{   } S_{ij} \equiv \braket{\hat{c}^{\dagger}_{i\uparrow}\hat{c}^{\dagger}_{j\downarrow}}_0,
\end{align}

\noindent
i.e., they represent averages of local hopping and pairing correlations in an uncorrelated state. $\lambda$ is variational parameter introduced to ensure that the wave function is normalized. The procedure of solving \eqref{var} is equivalent to diagonalization of the effective Hamiltonian

\begin{align}
     \mathcal{\hat{H}}^{eff} \equiv \sum_{ij\sigma} t_{ij}
     \hat{a}^{\dagger}_{i\sigma}\hat{a}_{j\sigma} +
     \sum_{ij} (\Delta_{ij}\hat{a}^{\dagger}_{i\uparrow}\hat{a}^{\dagger}_{j\downarrow} + \Delta^{*}\hat{a}_{i\downarrow}\hat{a}_{j\uparrow}),
\end{align}

\noindent
where

\begin{align}
    t_{ij}^{eff} \equiv \frac{\delta \mathcal{F}}{\delta P_{ij}}, 
\end{align}

\noindent
and

\begin{align}
    \Delta^{eff}_{ij} \equiv \frac{\delta \mathcal{F}}{\delta S_{ij}}. 
\end{align}

\noindent
In effect, the determination of the uncorrelated properties reduces to the diagonalization of the BCS--type Hamiltonian, and in turn, to that determining the properties in the correlated state. As the averages in that state are factorized in terms of uncorrelated $S_{ij}$ and $P_{ij}$, the latter procedure completes the determination of $\ket{\psi_G}$ provided the remaining variational parameters are also determined \cite{Spalek_phys_rep, Zegrodnik_2019_PRB}.

\subsection{Three-band model: A brief perspective}

As said above, strictly speaking, the elementary structural unit in two dimensions contains a single 3d$_{{x}^2-{y}^2}$ orbital due to the ninth electron of Cu$^{2+}$ ions and two 2p$_{x}$ and 2p$_{y}$ orbitals, each filled with two electrons in the parent (undoped) situation. Therefore, one has to formulate a three--band model to see at least, what is its connection to the widely used $t$-$J$-$U$-$V$ single--band models which should be regarded as a particular case of the present one. For that purpose, one starts from the Hamiltonian

\begin{align}
  \label{hammil}
    &\hat{\mathcal{H}} = \sum_{ij]'\alpha \sigma} t^{pp}_{ij}  \hat{p}_{i\alpha\sigma}^{\dagger}\hat{p}_{j\alpha\sigma} + \epsilon_{dp} \sum_{i\sigma} \hat{d}_{i\sigma}^{\dagger}\hat{d}_{i\sigma} \nonumber \\ &+ \sideset{}{'}\sum_{ij\alpha\sigma} t_{ij}^{pd}(\hat{d}^{\dagger}_{i\sigma}\hat{p}_{i\alpha\sigma} + \textrm{H.c.}) \\ \nonumber &+ 
    U_d \sum_i \hat{n}_{di\uparrow}\hat{n}_{di\downarrow} + U_p \sum_i \hat{n}_{pi\alpha\uparrow}\hat{n}_{pi\alpha\downarrow}.
\end{align}

\noindent
 \begin{figure*}
    \centering
    \includegraphics[width=0.9\textwidth]{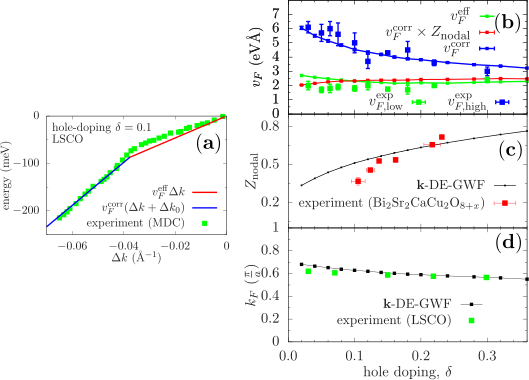}.
    \caption{(a) Experimental energy dispersion along the nodal direction for $\mathrm{La_{1.9}Sr_{0.1}CuO_4}$ extracted from Refs.~\cite{Zhou2003, Matsuyama2017}. The slopes of solid lines are obtained theoretically from the effective Hamiltonian (red) and first moment of the electron spectral function (red) for $\delta = 0.1$. (b) Doping-dependence of quasiparticle characteristic velocities above and below the kink (green and blue squares, respectively). Corresponding green and blue lines represent calculated effective- and correlated velocities calculated using $\mathbf{k}$-DE-GWF method. The red line is the correlated velocity multiplied by the calculated quasiparticle weight $Z_\mathbf{k}$. (c) Calculated $Z_\textbf{k}$ as a function of doping (black points and lines), compared with experimental data for $\mathrm{Bi_2Sr_2CaCu_2O_{8+\mathit{x}}}$ (red points, extracted from Ref.~\cite{Hashimoto2008}). (d) Calculated Fermi wave vector along the nodal direction compared with data for $\mathrm{La_{2-\mathit{x}}Sr_\mathit{x}CuO_4}$ for the same parameter values as those used in fitting Figures (a)-(c). After \cite{Fidrysiak_JPCM_2018}.} 
    \label{Fig4}
  \end{figure*}
In that model version the bare $p$-$p$ hopping is assumed as nonzero only between $nn$ 2p electrons, with  $t_{pp} \sim 0.5 \div 1 \text{ }eV$, $t_{pd} \simeq 1.1 \div 1.3 \text{ }eV$ is the single--particle hybridization between 3d and 2p states and induces an effective $d$-$d$ hopping in the higher order for the relevant antibonding states, $\epsilon_{d} - \epsilon_p \simeq 3.5 \text{ }eV$ is the so--called p $\rightarrow$ d charge transfer energy, whereas the relevant intraatomic $p$-$p$ and $d$-$d$ interactions have magnitudes $U_{pp} \simeq 4 \div 5 \text{ }eV$, and $U_{dd} \simeq 8 \div 1 \text{ }eV$. The other interactions term such as, e.g., that $\sim U_{pd}$ are neglected, what is probably an oversimplified feature of our model.

The question is to what extent results of the one--band model represented by \eqref{hamiltonian} and that starting from \eqref{hammil} are principally equivalent? A simple answer to this question is provided by inspection of the Fig. \ref{fig:fig2}b and noting that in the doped systems the interesting regime is then for $\delta \lesssim 1/3$. This, in effect, means that in the present version of the model the holes are located in the antibonding band, since the p--d charge transfer gap $\epsilon_{pd} = \epsilon_d-\epsilon_p$ is quite large on the scale of all parameters except $U_{dd}$. This bare--band picture persists also upon inclusion of interaction as the neglected $U_{pd}$ enlarges the charge transfer gap and $U_{dd}$ does not reverse the trend, when $n_{d} \lesssim 1$. The variational procedure presented in Sec.  3B is more involved \cite{Zegrodnik_2019_PRB} so it will not be presented in this minireview. Perhaps, it is worth showing explicitly the comparison of the doping dependence of the $d$-$d$ gap (cf. Fig. \ref{fig:fig2}(a)) in three--band model with that of single band model (cf. Fig. \ref{fig:fig2}b); the consecutive components $\Delta^{(i)}$ represent those between i-th neighbors (i-th coordination sphere). The component $\Delta^{(2)}$ is absent for the d--electrons, since the nearest neighboring $d$-$d$ correlations are strongly antiferromagnetic, inducing the effective spin--triplet correlations between the second neighbors. This exclusion does not appear for $d$-$p$ and $p$-$p$ pairing components; but they are of rather minor importance \cite{Zegrodnik_2019_PRB}. The two figures in the panel illustrate thus to what extent two models (1-- vs. 3--bands) may be regarded as equivalent . 

\subsection{The main qualitative features of the approach}

\begin{figure}
    \centering
    \includegraphics[width=0.5\textwidth]{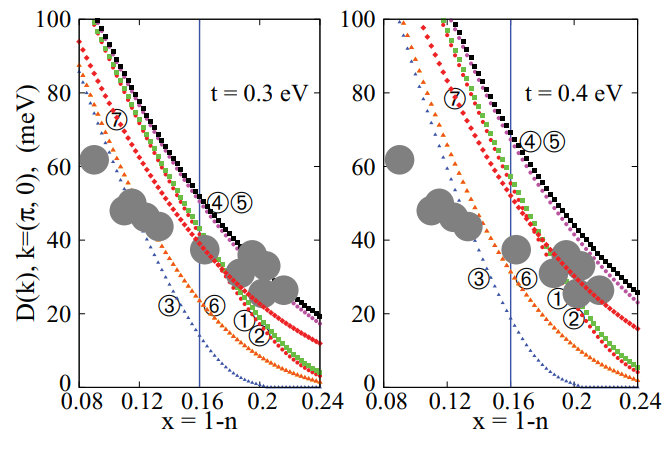}
    \caption{Doping dependencies of the SC gap $Delta_{\text{eff}}(\textbf{k})$ at $\textbf{k}=(\pi,0)$ for different approximation schemes \cite{Jedrak_PRB} 1--6 and for $t'/t=-0.27$ and $J/|t| = 0.3$. Large filled circles: experimental data. Two values of $t$ have been selected.}
    \label{Fig5}
\end{figure}

First of all, the two energy scales appear in a natural way, i.e., that described by $\ket{\psi_G}$ and $\ket{\psi_0}$, respectively, as exemplified by the physical gap $\Delta_G$ and pseudogap $\Delta_{eff}$. The question is whether those separate scales can be see in the actual correlated state. To illustrate the two faces of the correlated fermionic liquid we have compared first the doping dependence of the so--called pseudogap and the superconducting gap, $\Delta_{eff}$ and $\Delta_{dd}$, respectively. Those values have been compared with exemplary experimental results in Fig \ref{Fig3}ab. The amplitude $\Delta_{eff}$ is that obtained from the single--particle Hamiltonian \eqref{hammil}, whereas $\Delta_{dd} \equiv \Delta_G$ is that from solving the full expression $\braket{\mathcal{H}}_G$. Both gaps have d--wave symmetry $\Delta_{G, eff}(\textbf{k}) = \Delta_{G, \text{eff}}(\cos k_x - \cos k_y)$ from Fig. \ref{Fig5}ab. We see that the agreement of our theory (SGA) with experiment is rather qualitative, as only the trend of the data is reproduced. However, we believe that the inclusion of correlations (cf. Sec. V) may improve the results exhibit in Fig. \ref{Fig5}a essentially. Obviously, it is still to be carried out in the future (see also Fig. \ref{Fig3})b). 

As the second test of the two energy scales we consider the single--electron dispersion relation obtained from ARPES experiment. The exemplary comparison of our modified approach \cite{Fidrysiak_JPCM_2018} to experiment is shown in Fig. \ref{Fig5}b. We see that the comparison of the shifted by $\Delta k_0$ of the linear dispersion relation for the correlated particles (the part below the kink) with respect to that close to the Fermi energy. Note that $\Delta k \equiv k-k_F$ is the wave vector measured with respect to the Fermi surface point ($k=k_F$) in the nodal direction. In Fig. \ref{Fig4}b we exhibit the spectral two Fermi velocities $v^{eff}_{F,\text{low}}$ and $v^{corr}_{F}$, together with the renormalization factor $Z_{\text{nodal}}$ in the latter case (cf. the corresponding dependence in Fig. \ref{Fig4}c). Remaining labeling on those curves is self-explanatory \cite{Fidrysiak_JPCM_2018}.
\begin{figure}
    \centering
    \includegraphics[width=0.5\textwidth]{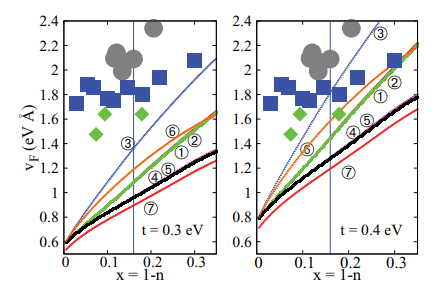}
    \caption{Doping dependence of Fermi velocity in the nodal [(0,0) $\rightarrow$ ($\pi$,$\pi$)] direction. Experimental data are marked by diamonds (after \cite{Jedrak_PRB}) YBa$_{2}$Cu$_{3}$O$_{{7}-\delta}$ (YBCO), squares (LSCO), and solid circles (BSCCO).Two $t$ values have been selected.}
    \label{Fig6}
\end{figure}
The basic question to ask is whether such a division into effective Landau quasiparticles and correlated particles of this quantum liquid (particles above and below the kink) is physically feasible. Our interpretation is that excitations the Fermi level (in the nodal direction) can be regarded as true quasiparticles in the Landau sense, albeit renormalized differently, since our starting interaction comprises \textbf{all} relevant itinerant electrons and is short--range and strong in real space. On the contrary, the single--electron excitations from the region deeper below the Fermi level (with energy $\Delta \epsilon \equiv |\epsilon-\epsilon_F| \gtrsim 0.1 \text{ }eV$) are dressed with the full interaction, in which the Hubbard term plays predominant role. Obviously, this division of a single quantum liquid of indistinguishable quantum particles into two parts is qualitative in nature and signals (by the kink's existence) a crossover behavior from a liquid of diluted quasiparticles to their truly correlated counterparts as one probes deeper into the Fermi sea. Such an interpretation requires a further test as it squares well with experiment (see also further evidence in the next Section). The division is coded in the selection of the wave function in the form \eqref{wav}, which contains a nonunitary projector $\hat{P}$, and is amplified by the fact that the starting (uncorrelated) wave function $\ket{\psi_0}$ is also of nontrivial nature and determined in a self--consistent manner that encompases also the states with broken symmetry from start. We should note at the end that such a mixed Fermi--non--Fermi liquid properties have been also observed in the transport properties \cite{Barisic}.

\section{Detailed testing of the theory: Equilibrium properties}

In this Section we discuss selected detailed characteristic of high temperature superconductors obtained within our real--space pairing among all itinerant electrons in our two--dimensional system.

\subsection{Inadequacy of the renormalized mean--field theory}
\begin{figure*}
    \centering
    \includegraphics[width=0.8\textwidth]{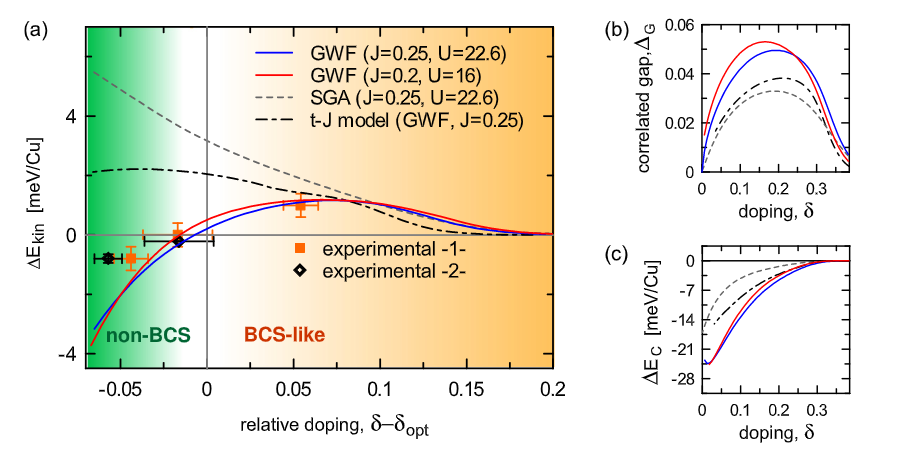}
    \caption{Selected superconducting properties: (a) Kinetic energy gain $\Delta E_{kin}$ vs. relative hole doping $\delta - \delta_{opt}$ ($\delta_{opt}$ is the optimal doping). The microscopic parameters are $J = 0.2|t|$, $U = 22.6|t|$ (for blue solid lines) and $J = 0.2|t|$, $U = 16|t|$ (for red solid lines); the experimental points are taken from Ref. \cite{Spalek_PRB_2017}.
    For comparison, the results obtained with SGA method (gray dashed line) and those for the t-J model ($J = 0.25|t|$) in DE--GWF approximation (dash--dotted line) are also included. Note that only the $t$-$J$-$U$ model solution describes the data in a quantitative manner. (b) correlated-gap magnitude $\Delta_G$ and (c) the condensation energy $\Delta E_C = E^{SC}_G - E^{PM}_G$, both vs. $\delta$, are also shown for the respective values of microscopic parameters and models. }
    \label{Fig7}
\end{figure*}

Our work started with the analysis of the so--called renormalized mean--field theory (RMFT), which has been very popular in the first decade after the discovery of superconductivity in the nonstochiometric oxide La$_{{2-x}}$Ba$_{{x}}$CuO$_{{4}-\delta}$ and  YBa$_{2}$Cu$_{3}$O$_{{7}-\delta}$. The approach was originally based on an improved version of the Gutzwiller approximation \cite{Anderson}. In our case, it take the form of statistically consistent Gutzwiller approximation (SGA) \cite{Jedrak_PRB}. In this approximation the regime of doping, where the superconductivity qualitatively as shown in Fig. \ref{Fig5}ab. The presence of antiferromagnetism at low doping can be reproduced qualitatively only after a careful selection of the detailed SGA approximation scheme is carried out \cite{Abram}. Furthermore, both the dependences of the (correlated) superconducting gap (cf. Fig \ref{Fig5}a) and particularly, of the dispersion relation of the single--particle excitations, obtained from ARPES (cf. Fig. \ref{Fig5}b) are not reproduced correctly. Explicitly, as we can see from the data included in Fig. \ref{Fig6}b and on the basis of our later analysis based on the full DE-GWF (cf. Fig. \ref{Fig4}), the Fermi velocity is rather flat, whereas the theoretical results shown in Fig. \ref{Fig6}b exhibit Fermi-liquid type of relative energy, diminishing steadily with decreasing doping. These results forced us to look for a theory, in which the SGA (or RMFT) results can be corrected is an essential way. In the next subsection we provide selected principal results illustrating the usefulness of our DE-GWF approach.

\subsection{Additional results: Beyond mean field theory and comparison with experiment}

The most striking result to a theorist may be the fact, discovered experimentally some time 
ago \cite{Deutscher2005}, is that the transition to the 
superconducting state, particularly in the regime of low doping, $\delta \lesssim 0.1$, takes place with the kinetic energy of the system getting lowered by the transition from the paramagnetic to superconducting phase. This is shown in Fig. \ref{Fig7}a, where the results (squares with the error marked) have been plotted against the relative doping 
$\delta-\delta_{\text{opt}}$, where $\delta_{\text{opt}}$ is optimal doping. Our theoretical 
curves require a more detailed explanation. Namely, the full curves represent our DE--GWF 
solutions for two slightly different values of parameters within $t$-$J$-$U$ model \cite{Zegrodnik_2019_PRB}. The 
other two (dashed and dot-dashed) curves represent the SGA and $t$-$J$ model (beyond--SGA) 
solutions, respectively. None of the latter two solutions reproduces the singular behavior at low doping, 
at least for the type of detailed approach chosen. Parenthetically, the fact that only the 
$t$-$J$-$U$ model, combined additionally with the DE--GWF, reflects the data in a quantitative 
manner, tells us that in order to reproduce fully them, one is forced to go beyond either
the Hubbard or $t$-$J$ model. In such a situation, we interpret the simultaneous presence of 
both the Hubbard term with realistic values of $U \sim 8-10 \text{ }eV$ and the kinetic exchange with its superexchange magnitude $J \sim 0.1 \text{ }eV$, as an implicit influence of the anionic 
2p$_{x,y}$ bands, not included in the standard one--band model, and producing the exchange interaction of desired magnitude, while keeping
the Hubbard $U$ in the realistic range at the same time.  

In Figs. \ref{Fig7}(b) and (c) we show the correlated gap magnitude of the d-wave solution and the condensation energy, respectively (the curve labelling and their meaning is the same as that in Fig. \ref{Fig7}a). Note that $\Delta_G \sim 0.03-0.05 = 15 \text{ }meV \sim 160 \text{ }K$ which is of the order of experimental value of $T_{{C}}$, but is substantially higher. This last fact is understandable as we do not account for thermodynamic fluctuation. Also, the condensation energy, i.e., the difference between the ground--state energies in normal and SC states is of the same magnitude and is strongly, but systematically, decreasing with increasing $\delta$. Comparing Figs. \ref{Fig7}(a) and (c) we see that surprising lowering
with diminishing $\delta$ is related to the corresponding kinetic--energy decrease. The lowering of $\Delta_G$ with $\delta \rightarrow 0$ is caused by the Mott--Hubbard localization effects (renormalization of $|t|$) so that the two quantities do behave differently near that limit. Such difference in behavior may be the sign of the quantum spin--liquid effects, which are interrupted by the carrier localization. For detailed discussion of phase diagram and associated with it crossover from non--BCS to BCS--like see \cite{Zegrodnik_2019_PRB}. 
\begin{figure}
    \centering
    \includegraphics[width=0.5\textwidth]{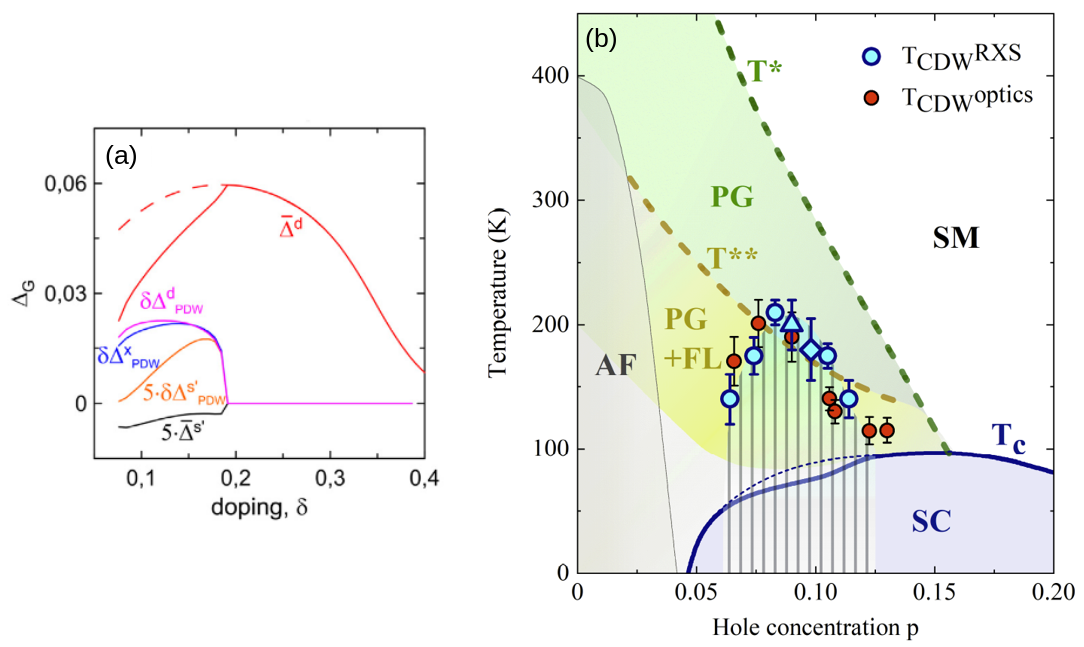}
    \caption{The phase diagram comprising various charge--density--wave states: (a) theory and (b) experiment \cite{Zegrodnik_2019_PRB, DiCastro2020}. For detailed discussion of various
order--parameter components see \cite{Zegrodnik_2019_PRB}. Note that the onset of pair--density wave (PDW) induces also a small s--wave type of ordering in the system
with the primary d--wave SC ordering. Pure d--wave superconducting phase appears only at and above the optimal doping, as observed.}
    \label{Fig8}
\end{figure}

In the last decade, the presence of the charge--order presence has been intensely discussed, also in the context of the appearance of hidden charge density--wave quantum critical point at the optimal doping \cite{Deutscher2005} . Leaving aside a detailed discussion, we have analyzed the effect of finite--range correlations within our DE--GWF method \cite{Zegrodnik_2019_PRB} on the appearance of the CDW--type state also with possible pair--density--wave presence have those states into the phase diagram. We plot in Fig. \ref{Fig8}a the theoretical results and have compared them with experimental data \cite{BialoPhD, Zegrodnik_2019_PRB}. Note the qualitative agreement between the two. In theory the most remarkable is the charge--splitting analog to the Fulde--Ferrell state in the split electronic structure which seems to reflect the experimental shape of the phase diagram. In passing, one can note the astonishing richness of the phases for this model two--dimensional structure. The relation of those features to the persistence of the van--Hove singularity in the correlated state, should perhaps be discussed in more detail.

\begin{figure}
    \centering
    \includegraphics[width=0.5\textwidth]{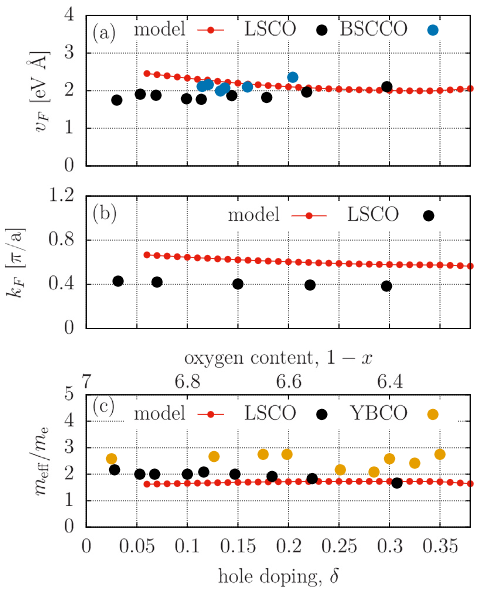}
    \caption{The basic characteristics calculated for the three-band model. From top to bottom: Fermi velocity $v_F$ , Fermi wave--vector $k_F$ , and effective
mass enhancement $m_{eff}/m_e$ ; all as a function of hole doping $\delta$. The parameters are: $t_{pd} = 1 \text{}eV$, $t_{pp} = 0.4 \text{}eV$, $\epsilon_{pd} = 3.2 \text{}eV$, $U_d = 11 \text{}eV$, and $U_p = 4.1 \text{}eV$. Note a quite smooth $\delta$ dependence of all the single--electron parameters.}
    \label{Fig9}
\end{figure}

To illustrate further the relevance of our results, we have calculated for the (approximately) same values of the microscopic parameters the selected single--particle characteristics in the correlated state and within the three--band model \cite{Zegrodnik_2019_PRB}. They are quite similar to those obtained within the single--band model \cite{Huffner_2008}. Explicitly, in Fig. \ref{Fig9} we display the doping dependence of the Fermi velocity (a), Fermi wave vector (b) and effective mass (c). What is surprising is a rather weak $\delta$ dependence in throughout the metallic phase. The dotted line represents out theoretical results, in the case (c) for two systems: La--Sr--Cu--O (LSCO) and Y--Ba--Cu--O (YBCO), respectively. Therefore, it is tempting to say that a nonmonotonic dependence of the critical temperature (T$_{{C}}$) or the correlated superconducting gap magnitude $\Delta_G$ is induced mainly by the competing character of the kinetic, exchange, and intraatomic Hubbard interactions. An analogical situation arises in the systems near the Mott--Hubbard insulator--metal transition \cite{SpalekPRL1987}. 
\begin{figure*}
    \centering
    \includegraphics[width=1\textwidth]{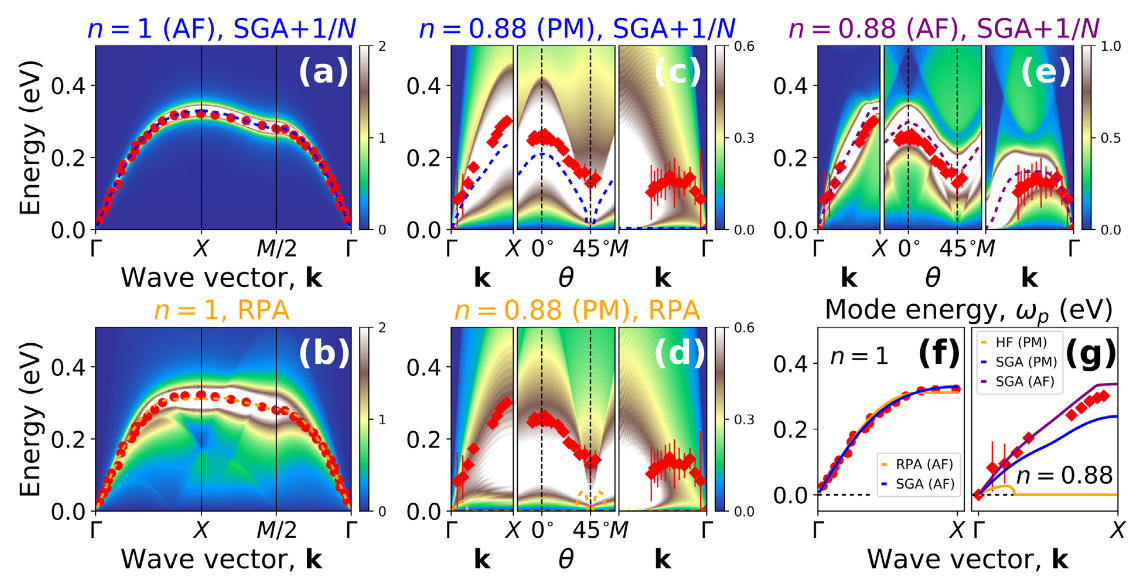}
    \caption{Imaginary parts of transverse dynamical spin susceptibility for obtained within one-orbital Hubbard model with nearest- and next-nearest-neighbor hopping integrals included, and comparison with experiment for $\mathrm{La_{2-\delta}Sr_{\delta}CuO_4}$ (LSCO). The model parameters are $t = -0.34\,\mathrm{eV}$, $U = 7|t|$. Panels (a) and (b) represent magnetic response at half filling ($n = 1$) in the antiferromagnetic state, obtained within $\mathrm{SGA}_x$+$1/\mathcal{N}_f$ and RPA, respectively. The spectra are similar and both of them match neutron scattering data for LSCO (red circles). The dashed lines are the paramagnon energies obtained from theoretical intensities using damped harmonic oscillator model. Panels (c) and (d) result from the same analysis, but for hole-doped system ($n = 0.88$) in the paramagnetic state. Here, the differences are qualitative: the $\mathrm{SGA}_x$+$1/\mathcal{N}_f$ method yields propagating magnetic excitations along $\Gamma$-$M$ line, whereas within RPA one obtains overdamped dynamics (see dashed curves and was discussion in the text). The agreement of the SGA+$1/N$ with RIXS data (diamonds) is semi-quantitative. Panel (e) shows $\mathrm{SGA}_x$+$1/\mathcal{N}_f$ results in the antiferromagnetic phase at lower temperature. In (f)-(g) we compare the theoretical RPA and $\mathrm{SGA}_x$+$1/\mathcal{N}_f$ paramagnon dispersion with experiment. Adapted from Ref. \cite{FidrysiakPRB2020}}
    \label{Fig10}
\end{figure*}

At the end, one should note that a discussion of the onset on nemacity appears also in the systems discussed here and 
discussed within DE-GWF \cite{FidrysiakPRB2020}. All in all, these results demonstrate the usefulness and effectiveness of the DE--GWF 
method
which represents a systematic approach beyond the mean--field type approach for these strong correlated systems. The whole approach 
bases on finite--U but large ($U\gtrsim W$) combined with strong superexchange interactions.

\section{Extension: Paramagnons and plasmons dynamic excitations}

So far, the whole DE--GWF analysis was based on taking into account, the static intersite correlations of increased range \cite{Spalek_phys_rep}, starting from SGA. We have extended this analysis to collective dynamic excitations (paramagnons and plasmons) by starting again from SGA and including long--range quantum fluctuations in the lowest order within $1/N$ expansion \cite{FidrysiakPRB2020, FidrysiakPRB2021, FidrysiakPRB2021_2, FidrysiakJPCM}. Here we summarize briefly only the results for spin fluctuations spectrum in the Gaussian approximation. The results are summarized in the panel composing Fig. \ref{Fig10}. The theoretical results marked by the color scale (on the right of the figures) and by the broken curves in figures (a)-(c). The curves in figures (f) and (g) are compared with the theoretical results explicitly with experiment, as well as show the differences for the doping $\delta=1-n=0.12$ between those obtained in random phase--approximation (RPA) and those obtained in SGA (i.e., without correlations included). Only the full theory SGA+1/N compares excellently with experiment. A similar theory can be formulated for the plasmon excitations \cite{FidrysiakPRB2020, FidrysiakPRB2021, FidrysiakPRB2021_2, FidrysiakJPCM} (see also the relevant contribution to this volume \cite{Fidrysiak_this}). One should mention that in describing the exhibited paramagnon excitation characteristics, a standard damped oscillator representation of the theoretical results was involved. This approximate (Lorentzian) representation of the excitation spectrum should be considered carefully, and perhaps, a more general approach is required. We should see a progress along this line in the near future. 

\section{Outlook}

We have overviewed here selected basic characteristics of high temperature superconducting cuprates. This paper summarizes some of the main results elaborated in detail in a comprehensive review \cite{Spalek_phys_rep}. The principal results and their favorable (semi)quantitative comparison with experiment support the fundamental concept of strong correlations combined with superexchange (kinetic exchange in one--band version of the theory) as the mechanism of the spin--singlet $d$-wave pairing in the cuprates. As we showed \cite{Spalek_phys_rep} and also here, it is indispensable to formulate the theory beyond any version of the (renormalized) mean field theory. What is still lacking is the incorporation of the quantum spin and charge fluctuations in the single--particle description of the normal--state properties to reproduce (or correct) the properties such as the linear electrical resistivity or pseudogap presence, to obtain a more complete quantitative picture. We should be able to see a progress along these lines in the near future. Also, the form of single--band starting Hamiltonian relation to its more--general three-band formulation with an explicit inclusion of the superexchange in the metallic phase should be reanalyzed carefully. Finally, the role of the third dimension is still to be incorporated on our theory to see explicitly the correction (if any) of the apical (interplane) oxygen may have in formation of the 3d superconducting state. 

Parenthetically, this article concludes the series of the minireviews published in the present journal over the years in this journal \cite{SpalekAPPA1,SpalekAPPA2}

\section*{Acknowledgement}

The work reported in this paper was financially supported by the grants OPUS Nos. ~UMO--2018/29/B\\/ST3/02646 and No.~UMO--2021/41/B/ST3/04070. The cooperaton and multiple discussions of the results wit Drs. Maciek Fidrysiak, Micha\l{} Zegrodnik, and Andrzej Biborski have been the most important factor in making this project so fruitful. I thank also to my Ph.D. student Maciek Hendzel for technical help during the writing of this report.

\section*{Appendix A: From classical Coulomb repulsion to extended Hubbard model}

The classical repulsive Coulomb interaction between two charges is long--range, changing with their mutual distance
$|\textbf{r}_i-\textbf{r}_j|$ as 

\begin{align}
    V_{12} \equiv V(\textbf{r}_i-\textbf{r}_j) = \frac{1}{\kappa} \frac{q_1q_2}{|\textbf{r}_i-\textbf{r}_j|}. 
    \label{A1}
\end{align}

\noindent
In the case of continuous charge densities $n(\textbf{r}_i)$ and $n(\textbf{r}_j)$ it takes the static Lenard-Wiechert form

\begin{align}
     V_{12} = \frac{e^2}{\kappa} \int d^3\textbf{r}d^3\textbf{r}' \frac{n(\textbf{r})n(\textbf{r}')}{|\textbf{r}-\textbf{r}'|}.
     \label{A2}
\end{align}

\noindent
For $n_i(\textbf{r}) = \delta(\textbf{r}-\textbf{r}_i)$~\eqref{A2} reduces to \eqref{A1}. In turn, in wave mechanics the interaction between two charges is then

\begin{align}
    V_{12} = \frac{e^2}{\kappa}\int d^3\textbf{r}d^3\textbf{r}' \frac{|\phi_1(\textbf{r})|^2|\phi_2(\textbf{r}')|^2}{|\textbf{r}-\textbf{r}'|}, 
    \label{A3}
\end{align}

\noindent
where now $|\phi_i(\textbf{r})|^2$ is the probability density for i-th particle. Finally, in 
quantum field theory the static interaction for indistinguishable particles is of the following 
operator form

\begin{align}
    \hat{V} = \frac{1}{2} \sum_{ijkl} \sum_{\sigma\sigma'} V_{ijkl} \hat{a}_{i\sigma}^{\dagger}\hat{a}_{j\sigma'}^{\dagger}\hat{a}_{l\sigma'}\hat{a}_{k\sigma}
    \label{A4}
\end{align}

\noindent
with 

\begin{align}
    V_{ijkl} = \int d^3\textbf{r}d^3\textbf{r}'\phi^{*}_{i\sigma}(\textbf{r})\phi^{*}_{j}(\textbf{r}')\frac{e^2}{\kappa|\textbf{r}-\textbf{r}'|} \phi_{k}(\textbf{r})\phi_{l}(\textbf{r}').
    \label{A5}
\end{align}

\noindent
Indices $(i,j,k,l)$ run over all possible single--particle states ${\phi_{i\sigma}(\textbf{r})}_{i=1,2,...,N}$ and $(\sigma,\sigma')$ are spin quantum numbers for particular fermions characterized by $(i,j,k,l)$. We see that in \eqref{A5} the probability densities $|\phi_i(\textbf{r})|^2$ and $|\phi_j(\textbf{r}')|^2$ are replaced by quantities
$\phi_i^*(\textbf{r})\phi_j(\textbf{r}')$ and $\phi_k(\textbf{r})\phi_l(\textbf{r}')$, respectively. We may say that they express roughly the overlap functions, but strictly speaking the states $\phi_i(\textbf{r})$ and $\phi_j(\textbf{r})$ are usually orthogonal, i.e. $\int d^3\textbf{r} \phi^{*}_i(\textbf{r})\phi^{*}_j(\textbf{r}) = \delta_{ij}$. 

The interesting us question is what happens if the wave functions ${\phi_i(\textbf{r})}$ are close to their atomic correspondants. In that limit $(i,j,k,l)$ when selected as parent atomic--state site positions are sufficiently far from each other that the largest contribution is to \eqref{A5}
comes from the term $i=j=k=l$ as in that case

\begin{align}
    V_{iiii} = \frac{e^2}{k} \int d^3 \textbf{r} d^3 \textbf{r}' \frac{|\phi_i(\textbf{r})|^2|\phi_j(\textbf{r}')|^2}{|\textbf{r}-\textbf{r}'|}, 
    \label{A6}
\end{align}

\noindent
i.e., has the \eqref{A3} from (effectively, also \eqref{A2} form). This is the reason why we call this term a director Coulomb interaction term. Then, \eqref{A4} in second--quantization reduces to the Hubbard interaction term if we take only \eqref{A6} out of all the terms appearing in \eqref{A5}, i.e., 

\begin{align}
    \hat{V}_{ii} \equiv \frac{1}{2} \sum_{i\sigma} V_{iiii} \hat{n}_{i\sigma}\hat{n}_{i\bar{\sigma}} \equiv U \hat{n}_{i\uparrow}\hat{n}_{i\downarrow}.
\end{align}

\noindent
In the other words, this term predominates over all remaining terms if the overlap functions $\phi^{*}_i(\textbf{r})\phi_j(\textbf{r}')$ are the only relevant quantities for $i=j$, i.e.,
the neighboring atomic states $\phi_i(\textbf{r})$ and $\phi_j(\textbf{r})$ are well separated. This assumption is the fundamental concept validating the Hubbard model which represents a particular limit of \eqref{A4}, namely

\begin{align}
    \hat{\mathcal{H}} = \sum_{ij\sigma} t_{ij} \hat{a}^{\dagger}_{i\sigma}\hat{a}_{j\sigma} + U\hat{n}_{i\uparrow}\hat{n}_{i\downarrow}.
    \label{A8}
\end{align}

\noindent
This (still unsolved) model applies to so many physical systems, albeit often only semiquantitatively. 

In the present analysis important are also nearest--neighbor Coulomb interactions. In that case, we have for two--state (two--site) terms namely

\begin{align}
    &\hat{V}_{} = \frac{1}{2} {\sum_{ij}}' K_{ij} \hat{n}_i \hat{n}_j - \frac{1}{2} {\sum_{ij}}' 
    J^H_{ij}(\hat{\textbf{S}}_i\hat{\textbf{S}}_j - \frac{1}{4}\hat{n}_i \hat{n}_j) \nonumber\\ &+ \frac{1}{2}
    {\sum_{ij\sigma}}' {V_{ij}}' (\hat{n}_{i\sigma}+\hat{n}_{j\sigma})(\hat{a}^{\dagger}_{i\bar{\sigma}}\hat{a}_{j\bar{\sigma}}+\hat{a}^{\dagger}_{j\bar{\sigma}}\hat{a}_{i\bar{\sigma}}) \\
    &+ {\sum_{ij}}'{J}'_{ij}(
    \hat{a}^{\dagger}_{i\uparrow}\hat{a}^{\dagger}_{i\downarrow}\hat{a}_{j\downarrow}\hat{a}_{j\uparrow}+ \text{H.c.}),\nonumber
\end{align}

\noindent
where the first term is the direct Coulomb intersite term, the second represents direct (Heisenberg) exchange interaction, the third the so--called correlated hopping, and the last pair hopping term (for details see \cite{Spalek_phys_rep}). The first two terms, when added to \eqref{A8} result in an extended Hubbard model. In the second term we include in the exchange (second) term containing also the effective kinetic exchange interaction. Hence, $J_{ij}$ becomes negative. In effect, by redefining constants $J_{ij} \equiv J^H_{ij}-J^{\text{kex}}$, $V_{ij} = K_{ij}/2 + J_{ij}/4$ we obtain t--U--J--V model in the form 

  \begin{align}
    \label{hamiltonianapp}
    &\hat{\mathcal{H}} = \sideset{}{'}\sum_{i\sigma} t_{ij} \hat{a}_{i\sigma}^\dagger \hat{a}_{i\sigma} + U \sum_i \hat{n}_{i\uparrow} \hat{n}_{i\downarrow} +  \sideset{}{'}\sum_{ij} J_{ij} \hat{\mathbf{S}}_i \hat{\mathbf{S}}_j  \\ \nonumber &+ \frac{1}{2}\sideset{}{'}\sum_{ij} V_{ij}\hat{n}_i \hat{n}_j.
  \end{align}

\noindent
Results obtained with this model and its particular versions are discussed in detail in the main 
text.

\bibliography{manuscript}

\end{document}